\documentclass[sigconf]{acmart}

\AtBeginDocument{%
  \providecommand\BibTeX{{%
    \normalfont B\kern-0.5em{\scshape i\kern-0.25em b}\kern-0.8em\TeX}}}

\setcopyright{acmcopyright}
\copyrightyear{2018}
\acmYear{2018}
\acmDOI{XXXXXXX.XXXXXXX}

\acmConference[Conference acronym 'XX]{Make sure to enter the correct
  conference title from your rights confirmation emai}{June 03--05,
  2018}{Woodstock, NY}
\acmPrice{15.00}
\acmISBN{978-1-4503-XXXX-X/18/06}

\usepackage{xr}
\usepackage{longtable}
\usepackage{xcolor,soul}
\soulregister\cite7
\soulregister\ref7

\usepackage{color, colortbl}
\definecolor{LRed}{rgb}{1,.8,.8}
\copyrightyear{2023}
\acmYear{2023}
\setcopyright{acmlicensed}\acmConference[CHI '23]{Proceedings of the 2023 CHI Conference on Human Factors in Computing Systems}{April 23--28, 2023}{Hamburg, Germany}
\acmBooktitle{Proceedings of the 2023 CHI Conference on Human Factors in Computing Systems (CHI '23), April 23--28, 2023, Hamburg, Germany}
\acmPrice{15.00}
\acmDOI{10.1145/3544548.3581074}
\acmISBN{978-1-4503-9421-5/23/04}

\begin{document}

\title{Cultural Differences in Friendship Network Behaviors: A Snapchat Case Study}

\author{Agrima Seth}
\email{agrima@umich.edu}
\orcid{0000-0001-8547-6304}
\affiliation{%
  \institution{School of Information, University of Michigan,}
  \city{Ann Arbor}
  \state{Michigan}
  \country{USA}
}

\author{Jiyin Cao}
\email{jiyincao@gmail.com }
\affiliation{%
  \institution{Stony Brook University}
  \city{Stony Brook}
  \state{ New York}
  \country{USA}}

\author{Xiaolin Shi}
\email{Xiaolin@snap.com}
\affiliation{%
  \institution{Snap Inc.}
  \city{Santa Monica}
  \state{California}
  \country{USA}
}

\author{Ron Dotsch}
\email{rdotsch@snap.com}
\affiliation{%
  \institution{Snap Inc.}
  \city{Santa Monica}
  \state{California}
  \country{USA}
}

\author{Yozen Liu}
\email{yliu2@snap.com}
\affiliation{%
  \institution{Snap Inc.}
  \city{Santa Monica}
  \state{California}
  \country{USA}
}

\author{Maarten W. Bos}
\email{maarten@snap.com}
\affiliation{%
  \institution{Snap Inc.}
  \city{Santa Monica}
  \state{California}
  \country{USA}
}

\renewcommand{\shortauthors}{Seth, et al.}

\begin{abstract}
Culture shapes people’s behavior, both online and offline. Surprisingly, there is sparse research on how cultural context affects network formation and content consumption on social media. We analyzed the friendship networks and dyadic relations between content producers and consumers across 73 countries through a cultural lens in a closed-network setting. Closed networks allow for intimate bonds and self-expression, providing a natural setting to study cultural differences in behavior. We studied three theoretical frameworks of culture - individualism, relational mobility, and tightness. We found that friendship networks formed across different cultures differ in egocentricity, meaning the connectedness between a user’s friends. Individualism, mobility, and looseness also significantly negatively impact how tie strength affects content consumption. Our findings show how culture affects social media behavior, and we outline how researchers can incorporate this in their work. Our work has implications for content recommendations and can improve content engagement.

\end{abstract}



\maketitle

\section{Introduction}
\label{sec:intro}
In the past two decades, social media platforms have transformed how individuals build and maintain their relationships. These platforms are increasingly becoming the preferred method for initiating intimate relationships \cite{whiteside2018helpful}, seeking advice \cite{pendse2019cross}, and community building \cite{pew}. With social media platforms becoming an integral part of social life for many of us (there are 4.26 billion social media users as of 2021 \cite{statista}), understanding the drivers of user behaviors is imperative.

Directly engaging with others (e.g., sending messages) and consuming their content (e.g., viewing, replying, and reacting to Stories and posts) are often studied to understand behavioral patterns on social media platforms. User behavior on online social media platforms can be said to be broadly driven by a complex combination of (a) user identity (personality, demographics), (b) the norms (descriptive and prescriptive) that the users in a network collectively subscribe to, (c) the relationship between users (friends, acquaintances, strangers), (d) usage intent; for example, professional (LinkedIn) vs. curated self-presentation (Instagram), and (e) platform affordances. While any particular platform usually provides the same affordances to all users on that platform, users bring their different backgrounds, experiences, expectations, beliefs, and values to the platform. As a result, different behaviors on the same platform are culturally influenced \cite{al2012impact,civelek2020differences,prabhakar2021toward,blachnio2016cultural}.

Most studies on social media user behavior are based on data that is west-centric \cite{yu_mccammon_ellison_langa_2016, sheldon2020culture}, and thus, their results have an implied context of western cultural norms. These findings fail to account for the heterogeneity in user behavior that arises from different cultural contexts \cite{pendse2019cross,alsaleh2019cross}. Hence, to further understand how cultural values affect behavior on these platforms, our work focuses on how users from different cultural backgrounds interact differently on a platform. Specifically, we use theoretical frameworks of cultural values to study the differences in the formation of friendship networks and the moderation of differential behavior of content consumption within these friendship networks. This paper uses three theoretical frameworks of cultural values: Hofstede’s concept of Individualism \cite{hofstede2001culture}, Thomson and colleagues’ concept of Relational Mobility \cite{thomson2018relational}, and Gelfand and colleagues’ concept of Tightness \cite{gelfand2011differences}. The data we used for our analyses is from the camera and messaging platform Snapchat. Snapchat is used in almost 150 countries and has 347 million daily active users worldwide \cite{snapdau}. Snapchat is a closed network, meaning that a lot of the content shared by individuals on Snapchat is only available to a limited set of trusted users. Past work on eliciting the motivations for Snapchat usage has shown that Snapchat is used to communicate with close relationships and is viewed as a platform with a relatively lower emphasis on self-presentation and impression management compared to platforms like Instagram \cite{alhabash2017tale,bayer2016sharing,piwek2016they,vaterlaus2016snapchat}. Because closed networks have less formal pressures and allow for intimate bonds and self-expression, they provide us with a cleaner setting to study differences in human behavior.

Specifically, we focus on 1) how culture influences network creation and 2) how culture influences content consumption behaviors embedded in the network. In particular, for the second question, we are interested in how culture moderates the effect of tie strength (i.e., the closeness between individuals) on content consumption. Past work has shown that tie strength strongly predicts a variety of user behaviors on platform, including what information will be exchanged \cite{yu_mccammon_ellison_langa_2016,warde2002social,panovich2012tie}, the likelihood to change one’s actions \cite{bond201261}, the attention given to content \cite{weng2018attention}, and the preferred behavior to signal engagement  \cite{arnaboldi2013egocentric}. We will explore how tie strength moderates tie strength's effect on content consumption. To study content consumption behavior, we use the metric of dwell time, i.e., the time a user spends consuming content that another user creates. 

In sum, we ask the following research questions:
\begin{enumerate}
\item How do friendship networks differ in countries with different cultural values?
\item How do cultural values change the effect of tie strength on dwell time?
\end{enumerate}

To answer our first research question, we studied the network properties of friendship networks across 73 countries, which have been surveyed by either \citet{hofstede2001culture}, \citet{thomson2018relational}, or \citet{gelfand2011differences}, and have different cultural values that lie on a continuum of the three cultural values of individualism, mobility, and tightness. We analyzed how friendship network size and egocentricity --- the extent to which a person’s friends are connected with each other --- vary across cultures in the closed network setting of Snapchat. We find that users from more individualistic, mobile, and loose cultures have a more extensive friendship network and are less egocentric. Next, we analyzed within these networks how tie-strength between users impacts engagement with content (dwell time) and the role of cultural values as a moderator. We found that individualism, mobility, and looseness negatively moderate the effect of tie strength on content consumption. 

Where previous work on culture and social media platforms has primarily been limited to a small sample size \cite{sheldon2020culture,al2012impact}, this paper contributes by studying cultural differences in user behavior on a large scale, analyzing hundreds of thousands of users across many countries. Further, where other quantitative works are usually limited to open or broadcast networks, this study explores relatively under-studied closed-network settings \cite{kaghazgaran2020social}.

From an HCI and design perspective, our work can advance our understanding of behavior patterns across cultures. We discuss the implications of understanding users’ engagement with content to design better experiences for the user. When applied to platform design, our work would help user-retention of platforms without compromising the user experience, in turn creating better outcomes for both users and platforms. Our work furthers the research that helps answer the question: What does it mean to understand and support users from diverse cultures on online platforms?  \cite{gallagher2013cross}. Most of the designs and practices of online platforms have a `one-size-fits-all’ approach and do not actively account for different user preferences across geographies. Our results provide evidence of differential behavioral patterns in online friendship networks across cultures and suggest how algorithm design can be culturally inclusive.

\subsection{Privacy and Ethics}
The data for this study was taken from Snapchat, and the study was conducted within Snapchat in accordance with Snapchat's policies and procedures with respect to Snapchat data. This analysis only uses the metadata of the user behavior. It does not analyze the actual content of the communication between the users.

\section{Related Work}
\label{sec:rw}
\subsection{Ties and user behavior}
Interpersonal relationships make social media platforms \textit{social}. Like in offline social networks, an individual's online network consists of individuals, with each of whom one shares a different type of relationship. Each dyadic relationship is different based on the closeness and the purpose they serve to the individual. Social network analysis literature uses the term \textit{tie strength} to differentiate between relations of different closeness. This term was coined by Granovetter\cite{granovetter1973strength}, who analyzed the role of different ties in different situations. The two types of ties characterized were \textit{strong} and \textit{weak}. The four dimensions determining a tie's strength were: the amount of time spent on a tie, the intimacy, the intensity, and reciprocal services \cite{granovetter1973strength}. Although researchers have used different operationalizations to conceptualize tie strength depending on the purpose of the study, many works on social media platforms operationalize tie strength as proportional to the total number of exchanges in the dyad. This operationalization of tie strength has been used to study various phenomena, like promoting mental well-being \cite{liu2022weak}, increased diffusion of information, and access to novel information \cite{granovetter1973strength, uzzi1999embeddedness}. While these works analyze the role of tie strength in reaping social benefits, studies have also focused on justifying Granovetter's hypothesis that the two ties elicit different interaction patterns, for which they analyze how information from different ties is received \cite{hershkovitz2020role, kramer2021strength, weng2018attention, kaghazgaran2020social}. These studies find evidence that individuals spend more time on the content received from stronger ties. 

\subsection{Cultural values}
One primary aspect of culture is that it's the normative value system that dictates acceptable practices and helps differentiate one group from another. Culture is both a result of the accepted past actions and the determinant of acceptable future actions. One of the ways to reason about attitudes and actions is to understand the culture people are in. Prior studies have shown that an individual’s behavior in the online space is influenced by their culture in the same way as offline behaviors.  With cultural values shaping actions, we must first understand how culture can be measured and then how culture affects behavior. While prior work usually focuses on groups and their specialized culture, we introduce literature from cultural psychology in our work. Culture is often operationalized through \textit{dimensions} where a \textit{dimension} is defined as “an aspect of a culture that can be measured relative to other cultures.” 

In this paper, we bring in concepts from three dominant cultural psychology theories, namely, individualism-collectivism \cite{hofstede2001culture}, relational mobility\cite{thomson2018relational}, and tightness-looseness\cite{gelfand2011differences}, to explore how culture impacts network creation and content consumption behaviors within a network. Below, we briefly introduce each of the cultural dimensions.

\subsubsection{Hofstede’s Individualism-collectivism}\citet{hofstede2001culture}  analyzed data from over 50 countries and identified six critical dimensions of national culture. Individualism-collectivism is one dimension that has drawn the most research attention. Typically, individualism leads to loose ties among the individuals of a society. Individualists focus on "I" as opposed to "we." Because groups are less important to them, individualists also tend to show no difference in their behaviors and attitudes toward ingroups versus outgroups.
In contrast, collectivism leads to a collective identity, and the welfare of an individual is implicitly assumed to be linked to the interests of the larger group. Hence, collectivists focus on "we." Because of their particular focus on "we," collectivists are known to have different norms and behaviors towards ingroups versus outgroups and place greater emphasis on harmony. 

Because of the "I" nature, individualists need to constantly reach out to build networks and also tend to see relationships as fluid. In contrast, collectivists see relationships as given, and thus, they are less active in building networks. As a result, we predict that individualism will be positively correlated with friendship network size. 

Individualists are less likely to treat other people based on relationship strength and group membership, whereas collectivists tend to have a strong tendency to favor ingroup members and people they are close to. This should also be manifested in how tie strength drives content engagement behavior in different cultures. As a result, we predict that individualism will negatively moderate the positive effect of tie strength on content engagement, such that the effect of tie strength on content engagement will be weaker for individualists than for collectivists.   

Hence, we hypothesize that:
\\
\textit{H1a: The friendship network for individualistic cultures is larger than the friendship network for collectivistic cultures}\\
\textit{H1b: Individualism negatively moderates the effect of tie strength on content engagement.}

\subsubsection{Relational Mobility} \citet{thomson2018relational} conducted a survey across 39 countries using a set of 12 questions to construct their dimension of culture. Relational mobility indicates  the degree of freedom and opportunities the members of a culture have to form and terminate relationships. The two opposing poles on this index are high and low relational mobility. For example, relational mobility is high in North America and low in Japan. Because relationships in high-mobility cultures are less stable and easier to change than those in low-mobility cultures, they are more fragile. It also requires more effort to maintain committed relationships. Prior work has shown that cultures with higher relational mobility tend to share more about themselves (self-disclosure), are more active in giving support, and tend to have more trust in the members of the society \cite{yuki2020psychological, thomson2018relational}. Because cultures high in mobility have more opportunities to form relationships, it allows individuals to have a larger network. In a similar vein, because in high mobility cultures, individuals see relationships as more fragile and fluid, they are less likely to adjust their interpersonal behaviors based on tie strength. As such, we predict that relational mobility will negatively moderate the effect of tie strength on content engagement, such that the effect of tie strength on content engagement will be weaker in high-mobility cultures than in low-mobility cultures. 

Hence, we hypothesize that:\\
\textit{H2a: The friendship network for high mobility cultures is larger than the friendship network of low mobility cultures}\\
\textit{H2b: Relational mobility negatively moderates the effect of tie strength on content engagement.}

\subsubsection{Tightness} \citet{gelfand2011differences} conducted a survey across 33 countries using 12 behaviors across 15 situations to construct their dimension. Tightness-looseness is about the extent to which a society tolerates norm-deviant behaviors. The two opposing poles on this index are tight and loose. For example, Looseness is high in North America and low in Japan. Tight cultures have stronger norms and are less tolerant of behavior that deviates from the norm. In contrast, loose cultures have relatively weaker norms and are more tolerant of behavior that deviates from the norm. As such, we predict that tightness should be negatively correlated with network size because a tight culture makes it hard for people to bring new members to a social network. Cultural tightness is often considered a selection criterion to test whether a new member can fit in. In contrast, the level of scrutiny will be much lower in a loose culture, making it easier for an individual to expand their network.
Similarly, we predict that tie strength's effect on content engagement will be weaker in loose cultures than in tight cultures. In a loose culture, tie strength is less likely to be seen as a criterion that individuals rely upon to decide how they approach a person. In contrast, in a tight culture, tie strength is a monitoring mechanism that powerfully regulates people. As a result, people draw more influence from tie strength, including content engagement behavior. 

Hence, we hypothesize that:
\\
\textit{H3a: The friendship networks for tighter cultures are smaller than friendship networks of looser cultures}\\
\textit{H3b: Tightness positively moderates the effect of tie strength on content engagement.}\\

Although the three cultural dimensions originated from different theories, they are often conceptually related. Prior work has shown that individualism, relational mobility, and looseness are often moderately correlated (Thomson et al., 2018, Appendix Table S8, p. 51 \cite{thomson2018relational}). For example, the U.S. is a culture that is individualistic, high mobility, and loose at the same time, whereas Japan is a culture that is collectivistic, low mobility, and tight. However, while Germany ranks higher in individualism and mobility, it ranks lower in looseness, whereas Brazil, though less individualistic, is more mobile and loose. Thus, while the three theories are conceptually related and can serve as a robustness check for one another, they each touch upon a unique cultural aspect. When researchers study the effect of one of the cultural values on individuals, they also tend to include the other two as a way of robustness check \cite{thomson2018relational,talhelm2014large}. As a result, although the three dimensions are from different theories, we see them as a whole package.

In sum, culture provides an important context about the shared common knowledge to its members on how to behave in a given context and how others will interpret their behavior. Comparative work on interpersonal relationships across cultures has shown that the same relationships elicit different behaviors in different cultures, implying that the same relationships across cultures are not similarly perceived  \cite{raeff2000conceptualizing,ting1996communication,goodwin2013personal, obal2016cross,gupta2018cultural,errasti2018differences}. Our work aims to analyze if user behavior on the same online platform provides empirical evidence that the impact of tie strength on their behavior varies across cultures.

\section{Data}
\label{sec:data}
We conduct our study on the Snapchat platform. Snapchat is an online messaging platform where content shared between users is ephemeral. Like most platforms, Snapchat allows users to exchange content in the form of text, images, and videos. The interactions between users can be one-to-one, one-to-group, or one-to-all friends (a broadcast interaction). Interactions are identified by different names and are introduced below:

\begin{itemize}
\item Snaps: A direct or personal interaction of image or video content type between users, which may be one-to-one or one-to-group. Depending on the receiver's chosen settings, Snaps disappear immediately after viewing or 24 hours later. In our analysis, we only consider Snaps that are exchanged between dyads (just two users), which are termed `direct Snaps.' We do not analyze Snaps sent to groups.
\item Chats: A text message between users. Akin to Snaps, depending on the receiver's chosen settings, chats disappear immediately after viewing or 24 hours later. In our analysis, we only consider the chats that are exchanged between dyads (just two users), which are termed 'direct chats.' We do not analyze group chats.
\item Stories: A broadcast interaction (with all of one's friends) having an image or video as the content type. Users on Snapchat (posters) can create Stories for their friends (viewers) to consume. Stories constitute a pull communication wherein friends decide to either engage with a Story in part or whole or ignore it. Unlike Snaps and chats that disappear after watching, Stories are available for 24 hrs after posting and can be viewed multiple times. 
\end{itemize}

We analyze users on Snapchat who share a friend connection. Friendships on Snapchat are bidirectional and are unlike the `follow model' that platforms like Instagram and Twitter allow (i.e., both individuals need to add each other as \textit{friends} in Snapchat). For each of the 73 countries (Refer appendix \ref{sec:countrylist}), we randomly sampled 10,000 unique users (egos), their associated Story viewing activity for one month, and their complete one-hop friend network. Though users may have friends across geographies, we filtered the data only to include those friend pairs where both friends resided in the same country. Aggregated over all 73 countries, cross-country friendships accounted for 21.8\% of the data. The filtering resulted in a total dataset of approx 600,000 users per country. Each user can view Stories from multiple friends, with each of whom they share a different level of closeness. This results in a data set of unique dyadic relations between a Story viewer and a Story poster. For each dyadic interaction, we calculate aggregated statistics of the total time spent by a viewer on each of the poster's Stories, the total number of Stories shared by a poster, and the total number of Snaps and chats exchanged between the two in the dyadic communication. To avoid noise from users who rarely engage with each other, we only keep those dyadic pairs where at least one direct chat or Snap has been exchanged by both the Story poster and the viewer during the one month we analyzed. To control for effects unrelated to the cultural values but caused by the economic development and platform reach in a country, we include each country's GDP \cite{imf}, which is a measure of a country's economic standing, GINI \cite{wb}, which is a measure of economic inequality within a nation, and Snap's market penetration \footnote{Internal Snap INC. marketing data}, which measures the user-base of Snapchat for a country. Section \ref{sec:method} details the process used to answer each research question. The three cross-cultural theories that inform our study did not survey all the same countries. Thus, while the three theories do not have a perfect overlap with each other (Refer appendix {\ref{sec:countrylist}}), using all three allows us to cover 73 unique countries.

\section{Method}
\label{sec:method}
We use the observational data from Section \ref{sec:data} and create statistical models to understand the role of culture on users' network formation and content engagement (dwell time). Building on and aligning with prior cross-cultural work, we consider a country a representative unit of one culture \cite{hofstede2001culture,thomson2018relational,gelfand2011differences} and analyze the users at the group level of a country.

\subsection{RQ1: How do friendship networks differ in countries with different cultural values?}
We first measured each country's average friendship network size to determine whether people from different cultures have  different friendship networks. For this, we calculated the total number of friends per user in each country and averaged it over the total number of users in the country. 

Next, for each country under study, we reconstruct the ego network (egonet) for that country’s randomly sampled 10,000 users. An ego network consists of the user (the ego), the user's friends (the alters), and the friendship relations between the alters. The egonets formed were independent, i.e., the users' egonets did not overlap. We filter out networks that consist of only two nodes (users who are only connected to the default Snapbot and do not have other friends on the platform) or star graphs (a pattern where a user is connected to other users, but none of those other users are connected, which is a pattern mainly shown by bots \cite{schuchard2018bots, uyheng2020bots}). Since all friendships on Snapchat are bidirectional, we convert the graph to a simple graph by removing the multiple edges (edges that are incident on the same pair of nodes). For each of the egonets, we calculate measures of egocentricity - the density, transitivity, and the betweenness centrality of the ego using the igraph package in R \cite{igraph}. 

Ego betweenness measures the percentage of shortest paths between two alters. In a social network setting, it allows us to measure the importance of the ego node. The higher the betweenness centrality, the more the ego node is the binding factor between its friends. Since centrality is sensitive to network size, we normalized it by the maximum possible betweenness of the ego node. This approach is in line with prior work on measuring betweenness in egonets Na et al.,\cite{na2015new}. 

\begin{displaymath}
\text{Betweenness centrality of node i}=\sum_{i\neq j \neq k} \frac{g_{jk} (i)}{g_{jk}}
\end{displaymath}
Where $g_{jk}$ is the number of shortest paths that connect node j and node k, $g_{jk} (i)$ is the number of these shortest paths that include node i.

Network density is the ratio of the edges in the user's network to the edges of the same user's hypothetical network where every node is connected to every other node. Likewise, transitivity is the number of triads relative to the number of possible triads. In our setting, density and transitivity measure the tendency of the users to cluster or connect. The higher the density and transitivity, the more the tendency of the group to cluster.

\begin{displaymath}
\text{Density for an undirected graph}=\frac{\sum_{j \neq k} z_{jk}}{\frac{n*(n-1)}{2}}
\end{displaymath}
Where n is the number of nodes in a network, and $z_{jk}$ is equal to 1 if the alters j and k are connected.

\begin{multline*}
    \text{Transitivity for an undirected graph} = \\ 3*\frac{\text{number of triangles in the network}}{\text{number of connected triples of nodes in the network}}
\end{multline*}

A high density and transitivity are indicative of people connecting with friends of friends; a low betweenness, on the other hand, implies a reduced tendency of nodes to cluster together. Prior work by \citet{na2015new} on self-reported Facebook networks in East Asia and the USA found that users from the USA were more egocentric than users from East Asia (had higher Ego Betweenness and lower Density and Transitivity). We use the same methodology --- to analyze data across more countries --- to explore whether these findings generalize across platforms and for data that is not self-reported but an individual's actual network data from a social media platform. To maintain consistency with Na et al.,\cite{na2015new}, we log-transform density and transitivity and then inverse the transformation by multiplying minus one; we transform betweenness using $log(1 + Max(x) - x)$ and then inverse the transformation by multiplying minus one.

\subsection{RQ2: How do cultural values change the effect of tie strength on dwell time?}

Online social media platforms continually aim to remove obstacles for content creation and consumption; this has allowed for a myriad of content to be available for consumption by users on all platforms. With the multitude of content available, attention from one’s social network has become a valuable and competitive resource. Here, we analyze how users allocate their attention to social connections with varying degrees of closeness and how this allocation is moderated by culture. We study attention in the context of Stories posted by friends in one’s network. We examine whether tie strength predicts one’s dwell time on a Story and whether culture moderates the relationship. 

\subsubsection{Measuring interest}
 Attention to a poster’s Story is a proxy for the interest in the information shared by the user. Attention towards a friend who posts Stories (p) is measured by the total time they spend on viewing their Story; longer attention (\textit{dwell time}) for a Story indicates a stronger interest towards that friend. To measure total time spent on content consumption (TC), we refer to the formulation proposed in prior works on measuring content \textit{dwell time} \cite{kaghazgaran2020social}. 
\begin{displaymath}
TC(v,p)=\sum_{s \in S_{p \rightarrow v}} \delta(s)
\end{displaymath}
where $S_{p \rightarrow v}$ denotes the set of Stories posted by p and consumed by v, s denotes (without loss of generality) one such Story sample, and $\delta(s)$ indicates the time spent by v in viewing the Story. This measures the relative difference in the viewer’s interest across different posters. However, as pointed out in prior literature, a viewer’s total view time on a poster’s Story can be skewed by the frequency of the posting activity of the Story creator, i.e., given the equal likelihood to consume Stories from different poster’s $TC(v,p1) > TC(v,p2)$ if $|S_{p1 \rightarrow v}| > |S_{p2 \rightarrow v}|$. Hence, we model \textit{dwell time} towards a sender \textit{s} as the average time spent by a viewer on the sender’s Stories. 
\begin{displaymath}
DT(v,p)= \frac{\sum_{s \in S_{p \rightarrow v}} \delta(s)}{|S_{p \rightarrow v}|}
\end{displaymath}
Dwell time is measured in seconds. While Stories vary in duration and can, in turn, influence dwell times, our initial analysis of viewing time distribution showed that most viewing activities were short and independent of content duration. This finding is in line with prior works on dwell time in closed network settings \cite{kaghazgaran2020social} - thus, we do not control for this variable.

\subsubsection{Measuring social tie strength between two users}
Tie strength between two users is a complex concept, subject to user perceptions and emotions; hence a direct quantitative measure of tie strength between users is challenging. However, measuring the activity of direct conversations between two users on social media platforms has proven to be an effective proxy in estimating tie strength: the higher the number of dyadic message exchanges, the closer the two users are. Some users send burst messages while others send fewer but longer messages; thus, we model tie strength (TS) as the total number of direct Snaps and chats exchanged between a pair of users.
\begin{displaymath}
TS(v,p)= |DC_{p \rightarrow v}| + |DC_{v \rightarrow p}| + |DS_{p \rightarrow v}|+|DS_{v \rightarrow p}|
\end{displaymath}
where $DC_{p \rightarrow v}$ denotes the set of direct chats sent by the Story poster to the Story viewer, $DC_{v \rightarrow p}$ denotes the set of direct chats sent by the Story viewer to the Story poster, $DS_{p \rightarrow v}$ denotes the set of direct Snaps sent by the Story poster to the viewer, and $DS_{v \rightarrow p}$ denotes the set of direct Snaps sent by Story viewer to the poster. Preliminary analysis of tie strength in each country showed variation; hence, we standardize tie strengths within each country and use the standardized version for analysis. 

\subsubsection{Measuring culture of each user}
We use the results from Hofstede’s Individualism \cite{hofstede2001culture}, Thomson et al.’s Relational Mobility \cite{thomson2018relational}, and Gelfand et al.’s Tightness \cite{gelfand2011differences} dimensions, discussed in Section \ref{sec:rw} as the measure of cultural values (CV) for the country that an individual belongs to. These measures have been widely used in the literature. Hofstede’s work has attracted over 45,000 citations, Thomson et al.’s (more recent) work has already been cited 178 times, and  Gelfand et al.’s work has more than 2000 citations. Since each value system is on a different scale (Appendix \ref{sec:countrylist})--- Individualism ranges from 6 to 91, Relational Mobility ranges from 3.886 to 4.607, and Tightness ranges from 1.6 to 12.3 --- we independently standardize each value system across countries and use the standardized version for analysis.

\subsubsection{Mixed effects model to analyze dwell time as a function of tie strength and cultural values }
We used a linear mixed-effects model to address the research question of how cultural values moderate the impact of tie strength on the time spent consuming content (dwell time) in closed network settings. Since the sets of countries surveyed by \citet{hofstede2001culture}, \citet{thomson2018relational}, and \citet{gelfand2011differences} do not have perfect overlap, we created three multilevel models to understand how cultural values moderate the effect of tie strength on Story dwell time. The models included terms for tie strength (dyad level), cultural value (country level), and their interaction as fixed effects, with random intercepts for country and viewer, and the number of friends, the GDP, GINI, and Snap’s market penetration (MP) for a country as control variables. We standardized each value system across countries and used the standardized version for analysis. Since we have multiple observations per country and a viewer views multiple posters, we include the random effects due to the country and the viewer. 

\begin{multline*}
DT (v,p) =  TS (v,p)\ X\ CV (v) + |v_f| + GDP + GINI + MP + \\
(1 | country) +  (1 | Viewer)
\end{multline*}
where $|v_f|$  refers to the number of friends a viewer has, $TS (v,p)$ is the tie strength between a pair of viewers and a poster, and $CV(v)$ is the cultural value of the viewer, which is the same as the cultural value of the poster.

Since each dyad contains the dwell time of multiple Stories, we model random effects for the dyad. However, users in a dyad can have two roles: sometimes a user is a viewer, and sometimes a poster. A user who is a viewer (v) for a poster p can be a poster ($p'$) for some other node ($v'$). This directionality complicates modeling. To simplify, we randomly regard one person as the viewer and the other as a poster, disregarding the Stories of that dyad where the viewer posted and the poster viewed. To ensure that the results \ref{sec:results} are robust against role assignment, we bootstrapped the analysis; on each run, for each dyad, viewer and poster roles were randomly assigned before fitting the model. The bootstrapped results are in Appendix \ref{sec:bootsrap}.

\section{Results}
\label{sec:results}
\subsection{RQ1: How do friendship networks differ in countries with different cultural values?}
We report zero-order Pearson correlations between cultural values and friendship network size in Table \ref{tab:friendshipsize}. We find that countries that rank higher in individualism, mobility, and looseness tend to have a bigger friendship network than collectivistic, less mobile, and tighter countries. This means that people in the higher ranking countries are connected to more friends on Snapchat, supporting \textit{H1a, H2a, and H3a}. To check for robustness, we ran the same analyses with GDP, GINI, and Snapchat's market penetration as control variables. The addition of control variables reduced the sample size of countries, but the results corroborate those reported here \ref{sec:friend_control}.  

Next, the structural analysis of the ego networks of users from different cultures (Table \ref{tab:egonet}) shows that the ego centrality of user networks on Snapchat varies with cultural values. Akin to Na et al.,\cite{na2015new}, we find that the individual structural measures, namely density, transitivity, and betweenness, are highly correlated (Table \ref{tab:corr}), and thus we average the standardized values and report the results for this averaged index of ego-centrality. The results show that mobility and individualism are negatively correlated with egocentricity, and tightness is positively correlated with egocentrality. This means that in countries that rank higher on mobility and individualism, people's friends on Snapchat are more likely to be connected to each other, and in countries that rank higher on tightness, people's friends on Snapchat are less likely to be connected to each other.

\begin{table} [h]
  \caption{Pearson correlation between cultural values and friendship network size $(* p<0.05, ** p<0.01, *** p<0.001)$ }
  \label{tab:friendshipsize}
  \begin{tabular}{|c|c|c|}
    \toprule
    Cultural Value & Correlation & Number of countries\\
    \midrule
    Individualism & 0.68** & 65\\
    Relational Mobility & 0.31* & 37\\
    Tightness & -0.37* & 30 \\
  \bottomrule
\end{tabular}
\end{table}
\vspace{-0.5cm}


\begin{table} [h]
  \caption{Pearson correlation between network structural measures for data across different cultural values after controlling for GDP, GINI, and market penetration $(* p<0.05, ** p<0.01, *** p<0.001)$}
\label{tab:corr}
 \begin{tabular}{|p{1.7cm} |p{2.5 cm} | p{2cm} | p{2cm}|}
  \toprule
   Cultural Value & Betweenness and  Transitivity & Betweenness and Density & Density and  Transitivity\\
   \midrule
    Individualism & 0.74*** & 0.504*** & 0.92 ***\\
    Mobility &  0.76 *** & 0.49** & 0.85***\\
    Tightness &0.82*** & 0.45* & 0.83 ***\\
    \bottomrule
  \end{tabular}
\end{table}

\begin{table} [H]
  \caption{Pearson correlation between cultural values and egocentrality$(* p<0.05, ** p<0.01, *** p<0.001)$ }
  \label{tab:egonet}
  \begin{tabular}{|c|c|}
    \toprule
    Cultural Value & averaged index of ego-centrality \\
    \midrule
    Individualism & -0.07 ***\\
    Relational Mobility & -0.04***  \\
    Tightness & 0.06***\\
  \bottomrule
\end{tabular}
\end{table}


\subsection{RQ2: How do cultural values change the effect of tie strength on dwell time?}
Given that the friendship network structures are different across cultures, using multilevel modeling, we analyzed how cultural values moderate the effect of tie strength on the viewer's dwell time (Tables \ref{tab:idv}, \ref{tab:mobility}, \ref{tab:tightness}). We see that an increase in the strength of ties increases the dwell time, a result in line with prior works \cite{kaghazgaran2020social,weng2018attention}. Having more friends reduces a viewer's dwell time on content, which is likely because an increase in the number of friends leads to more potential Story content to consume. Though the cultural values do not have a significant main effect, they significantly moderate the effect of tie strength on dwell time across all three cultural values. We find that tie strength negatively moderates the effect of tie strength for more individualistic, mobile, and looser cultures. Thus confirming \textit{H1b, H2b, and H3b}. The bootstrap results from 100 runs corroborate the findings reported here in Appendix \ref{sec:bootsrap}. Our work focuses on understanding (and not predicting) within-dyad level dwell time from theories of country-level cultural values, which may not fully account for a lot of individual-level variation. However, a significant moderation effect allows us to argue for a substantiative effect of cultural values on individual-level behavior {\cite{moksony1990small}}. Using only the intersection of countries present across all three measures of culture, we check for robustness of these results (Appendix \ref{sec:culture_intersection}), and the results corroborate the results reported in Tables \ref{tab:idv}, \ref{tab:mobility}, \ref{tab:tightness}. Because the effects we found are on the smaller side, there is still a lot of unexplained variance, and we can not fully account for all individual-level and item (Story) level variation.


\begin{table}[h]
  \caption{Coefficients from Multilevel Modeling for the effect of Individualism as a moderator on Dwell Time $(* p<0.05, ** p<0.01, *** p<0.001)$, Sample size: country = 47, dyads = 460000, RMSE = 4.9, AIC = 2793115, BIC = 279226, R\textsuperscript{2} conditional = 0.04, R\textsuperscript{2} marginal = 0.01 }
  \label{tab:idv}
  \begin{tabular}{|c|c|c|}
    \toprule
    Fixed Effects& Estimate & Standard Error\\
    \midrule
    Intercept & 3.741*** & 0.078 \\
    Strength of Ties & 0.092*** & 0.007 \\
    Individualism & 0.035 & 0.074 \\
    Strength of Ties :  Individualism & -0.014*** & 0.007\\
    \hline
    \multicolumn{3}{|c|}{Control variables}\\
    \hline
    Number of Friends &  -0.338*** & 0.008\\
    GDP & -0.036 & 0.068 \\
    GINI & -0.040 & 0.060 \\
    Market Penetration &  0.065* & 0.059\\
  \bottomrule
\end{tabular}
\end{table}

\begin{table}[t]
  \caption{Coefficients From Multilevel Modeling for the effect of Mobility as a moderator on Dwell Time $(* p<0.05, ** p<0.01, *** p<0.001)$ Sample size: country = 26, dyads = 128800, RMSE= 3.12, AIC = 1438399, BIC = 1438504, R\textsuperscript{2} conditional = 0.27, R\textsuperscript{2} marginal = 0.01 }
  \label{tab:mobility}
  \begin{tabular}{|c|c|c|}
    \toprule
    Fixed Effects& Estimate& Standard Error\\
    \midrule
    Intercept & 3.835 *** & 0.097 \\
    Strength of Ties & 0.116***& 0.008\\
    High Mobility & 0.092& 0.071\\
    Strength of Ties :  High Mobility & -0.012*& 0.006\\
    \hline
    \multicolumn{3}{|c|}{Control variables}\\
    \hline
    Number of Friends &   -0.35*** & 0.011\\
    GDP &  -0.051 & 0.108 \\
    GINI & -0.02 & 0.102\\
    Market Penetration &   0.108 & 0.091\\
  \bottomrule
\end{tabular}
\end{table}
\begin{table}[H]
  \caption{Coefficients From Multilevel Modeling for the effect of Tightness as a moderator on Dwell Time $(* p<0.05, ** p<0.01, *** p<0.001)$, Sample size: country = 25, dyads = 100000, RMSE=2.19, AIC = 731754.3, BIC = 731850.8, R\textsuperscript{2} conditional = 0.80, R\textsuperscript{2} marginal = 0.01}
  \label{tab:tightness}
  \begin{tabular}{|c|c|c|}
    \toprule
    Fixed Effects& Estimate & Standard Error\\
    \midrule
    Intercept & 3.725*** & 0.1 \\
    Strength of Ties & 0.129*** &0.010\\
    Tightness & -0.060 & -0.082 \\
    Strength of Ties :  Tightness & 0.058***& 0.010 \\
    \hline
    \multicolumn{3}{|c|}{Control variables}\\
    \hline
    Number of Friends &  -0.283 *** & 0.011\\
    GDP &  -0.154* & 0.077\\
    GINI & -0.171 & 0.069\\
    Market Penetration &  0.179* & 0.077\\
  \bottomrule
\end{tabular}
\end{table}

\section{Discussion}
\label{sec:discussion}
Most social media platforms were introduced in the Global North before they started gaining a user base in other countries. As a result, studies on understanding users on social media platforms primarily draw from west-centric populations, which leads to unintended biases. Using data from 10,000 users per country from nearly 73 countries, our work studied how individuals across cultures differ in their behavior on the same platform. We control for confounders like the platform's market penetration, countries' GDP, and GINI score, which  may have influenced the platform's user base size and composition. Our main findings are:

\subsubsection*{Structure of friendship network} The analysis of the egocentrality of the friendship networks showed that individualistic, more mobile, and looser cultures are negatively correlated with egocentrality. This result is unlike the prior survey-based network analysis by Na et al. \cite{na2015new}, which found that individualism is positively correlated with ego centrality. Na et al. \cite{na2015new} recruited individuals through a call for survey participants on the Facebook platform, which resulted in a substantially varied number of respondents from each country and thus could be sensitive to selection and conformity bias. In our study, we randomly sampled users and analyzed the metadata of the user behavior, which provides a relatively cleaner signal for a user's choices. Apart from a more balanced number of users from different countries, we also analyzed data from a substantially higher number of countries. Apart from data collection and sample size differences, another potential source for the differences in findings could arise from who is befriended on these platforms.

Adams and Plaut posited that friendship's meaning varies substantially across cultures \cite{adams2003cultural}. Markus and Kitayama \cite{markus1991culture} argued that familial ties form an important part of a user's social network in collectivist cultures compared to individualistic cultures. With the demographics on Snapchat skewing towards a younger population \cite{demo1,demo2} and motivations differing from Facebook \cite{alhabash2017tale,vaterlaus2016snapchat,phua2017uses}, it is plausible that (a) the 'younger users' do not 'friend' familial ties due to the difference in how they make sense of 'friendship' and whom they 'friend,' and (b) the 'elder' familial members are absent from the platform. Since family ties form an important part of collectivist cultures, not including them on their Snapchat friendship network could be the reason for differences in our findings when compared to Na et al. \cite{na2015new}. While our results differ from Na et al., \cite{na2015new}, they agree with the findings from Igarashi et al.\cite{igarashi2008culture} that user's from collectivist cultures had more egocentric networks. Given that very few studies have explored how culture affects network structures, future work in this domain will help establish a stronger understanding of how culture influences the network structures formed on social media platforms.




Our findings bear important implications for future work that aims to study user interaction patterns on a platform. Firstly, studies should elicit and validate the network structure formed for their population of interest because the network structures vary across subpopulations on the same platform and across platforms, and relying on metrics from prior work with a mismatched population might lead to incorrect inferences. Next, the differences in friendship networks bear importance for context-aware friendship recommendation engines, which we discuss under design implications.

\subsubsection*{Cultural Values and user behavior} Culture is a complex societal-level phenomenon that guides individual behavior. Various studies have tried to study culture through a system of 'cultural values.' In this project, we chose three dominant theories in cultural psychology, ranging from Hofstede's dimensions published in 2001 \cite{hofstede2001culture} to more recent theories on Tightness and Mobility published in 2011 and 2018 \cite{gelfand2011differences,thomson2018relational}, respectively. Consistent with our hypothesis, we found that each cultural value (i.e., individualism, looseness, mobility) significantly moderates the effect of tie strength on dwell time, highlighting the significance of considering culture in understanding behavior patterns on social media. In addition, we found that individualism, looseness, and mobility moderates the relationship between tie strength and dwell time in the same direction. Theoretically, it is logical because in societies where people have more freedom to make friends and move between different circles (i.e., high relational mobility), a looser norm (i.e., looseness) is likely to develop, and a comparatively more self-focused mindset (i.e., individualism) is likely to rise. Indeed, prior work has also predicted that these three variables would have a similar impact on individual cognition and behavior \cite{thomson2018relational}. Thus, we extend the prior work in cultural psychology by adopting a cultural lens in understanding user behaviors on social media.

\subsection{Design Implications}
The diversity of content on platforms has made good recommendation systems a necessity. While these recommendation systems are becoming increasingly personalized, they fail to distinguish the varied meanings that different types of social ties have for users from different cultures. For example, if we consider the dyadic pair of user A, their strongest tie, and user B, their strongest tie, such that user A and B belong to different cultures, the influence of the respective strongest tie may be different. Our study, through evidence, argues for treating users and their friendship relations from different cultures differently when designing recommendation systems. Analyzing users at a cultural level may reduce the complexity of recommendation systems and make the recommendation system more culturally sensitive. By doing so, they may be able to better rank the content the user is more likely to engage with at a reduced cost. For example, our result suggests that when designing recommendation systems, tie strength should be given greater weight for users in less mobile, tighter, and collectivistic countries because our results show that tie strength is more strongly correlated to content dwell time in these countries.

Friendship recommendation engines that are unaware of 'how' and 'why' network structures differ across cultures run the risk of treating friending activities across different cultures as the same, resulting in a suboptimal  platform experience. For instance, the motivations of individuals from tight cultures could differ from those from loose cultures, i.e., in contrast to individuals from loose cultures, individuals in tight cultures might feel forced to friend not only those whom they want to but also those whom they have to  - say befriending familial ties. A recommendation engine that captures behavior from loose cultures might not be able to recommend users with whom one shares common friends. Similarly, a recommendation engine that focuses on tight cultures would explore less and over-recommend users with whom one shares common friends. Hence, using the behavioral understanding from only either of the cultures risks the failure of the algorithms ( and, in turn, platform experience) in the other cultures. Thus, while our work takes a step in highlighting 'how' the network structures differ, future work that provides insights into 'why' the network structures differ can further enrich the understanding of designing friendship recommendation algorithms.

\section{Limitations}
Our study is subject to a few important limitations.
First, our work uses data from Snapchat, which encompasses a significant but limited amount of people's online communications. We could only use available data for our study, and some of Snapchat's user data is only available for a limited time. Additionally, the actual content of Snapchat communications is not available for analysis. The Snapchat user group skews young \cite{demo2}, and studies have found that younger people have shifted away from traditional values \cite{sun2010value,nguyen2015using}. Second, recommendation algorithms play an important role in network formation on the platform. We did not have access to the friend recommendation algorithm for this study, and we could, therefore, not control for any potential confounding effects. Getting an insight into the algorithm and its impact on users across geographies could further enrich future work. Further, the focus of this study was to understand the friendship network and behavior on the online social network, which may differ from an individual's offline friendship networks and their interactions on these networks.
Next, not every country has been equally surveyed in prior research on cultural values. There is a non-perfect overlap between countries that have been studied for mobility and countries that have been studied for tightness. Once data from more countries becomes available, our analyses could be extended to include those countries. Future work can further build on ours by analyzing how content type interacts with cultural values and impacts dwell time. By using a large random sample of users across countries, country-level measures of economic growth, and inequity, we tried to limit selection bias and account for variations across countries. GDP and GINI measures help us control for country-level socioeconomic status. However, it is plausible that a given stratum of society is overrepresented on the platform, and country-level socioeconomic measures might not fully control for the platform user's socioeconomic status. The lack of finer-grained measures could be a limitation of the study.


Human behavior is complex and subject to factors that have individual-level variation. Hence, it is difficult to fully predict human behavior in the social sciences. The focus of our work was to test the theory of the effect of culture, as measured at the country level, on individual behavior. Like prior works, we can not fully account for all individual-level and item (Story) level variation. As brought out in the Introduction {\ref{sec:intro}}, individual behavior is affected by a host of other variables, and content engagement is no different. For example, the Story's content might be an important factor; however, we could not study this due to Snap Inc.'s policies on not retaining information about the content. Future studies can help make the model more complete by operationalizing the type of content and other variables that might affect the dwell time on content. While the cultural theories used in this study span a large geographic region, the identities of the researchers who created these measures could be a source of bias for these measures. As argued by Shweder {\cite{shweder1993cultural}} (p. 409), these studies can largely benefit from a more emic expansion approach, which would help remove biases from future empirical studies.




\section{Conclusion}
We examined the friendship network and the dwell time behavior of users across 73 cultures on the online platform Snapchat. We studied one month's data from 10K users from each culture. First, we found that the friendship networks curated by individuals from different cultures vary in size and egocentricity. We found evidence that individuals from individualistic, high mobility, and loose cultures tend to form larger friendship networks. We analyzed how cultural values moderate the relation between tie strength and users' content engagement behavior. We found that individualism, high mobility, and looseness negatively moderate this effect. This provides evidence for psychological theories which posit that relationships are not perceived similarly across different cultures, and thus their effect on user behavior is not uniform across cultures. Our work could advance the understanding of engagement with content on online platforms and how using this insight can improve recommendation systems. Incorporating cultural values in the experience design can improve the user experience and does better justice to the diverse backgrounds of platform users.

\begin{acks}
We thank Dr. Neil Shah (Snap Inc.) for providing valuable advice on friendship network creation and analysis.
\end{acks}

\bibliographystyle{ACM-Reference-Format}
\bibliography{sample-base}

\appendix
\section{Cultural value and friend network size with control variables}
\label{sec:friend_control}
\begin{table}[h] 
  \caption{Pearson correlation between cultural values and friendship network size with GDP, GINI, and Market Penetration as control variables $(* p<0.05, ** p<0.01, *** p<0.001)$ }
\label{tab:friendshipsize_appendix}
  \begin{tabular}{|c|c|c|}
    \toprule
    Cultural Value & Correlation & Number of countries\\
    \midrule
    Individualism & 0.6** & 47 \\
    Relational Mobility & 0.27 & 26\\
    Tightness & -0.51* & 24 \\
  \bottomrule
\end{tabular}
\end{table}

\section{Bootstapped results for Mixed effects model (across 100 runs)}
\label{sec:bootsrap}
\begin{table}[H]
  \caption{Bootstrapped Coefficients From Multilevel Modeling for the effect of Individualism as a moderator on Dwell Time }
  \label{tab:idv_appendix}
  \begin{tabular}{|c|c|c|}
    \toprule
    Fixed Effects& Estimate & $CI(95\%)$\\
    \midrule
    Intercept & 3.740 & [3.718,3.762] \\
    Strength of Ties & 0.103 & [0.100,0.106]\\
    Individualism & 0.033 & [0.028,0.038]\\
    Strength of Ties :  Individualism & -0.008 & [-0.011,-0.005] \\
    \hline
    \multicolumn{3}{|c|}{Control variables}\\
    \hline
    Number of Friends &  -0.329 & [-0.341,-0.317]\\
    GDP & -0.031 & [-0.038,-0.024]\\
    GINI & -0.040 & [-0.042,-0.039] \\
    Market Penetration &  0.56 & [0.038,0.074]\\
  \bottomrule
\end{tabular}
\end{table}
\begin{table}[H]
  \caption{Bootstrapped Coefficients From Multilevel Modeling for the effect of Mobility as a moderator on Dwell Time}
  \label{tab:mobility_appendix}
  \begin{tabular}{|c|c|c|}
    \toprule
    Fixed Effects& Estimate & $CI(95\%)$\\
    \midrule
    Intercept & 3.820 &[3.801,3.839] \\
    Strength of Ties & 0.114&[0.111,0.117] \\
    High Mobility & 0.092& [0.089,0.095]\\
    Strength of Ties :  High Mobility & -0.010& [-0.014,-0.007]\\
    \hline
    \multicolumn{3}{|c|}{Control variables}\\
    \hline
    Number of Friends &   -0.347& [-0.356,-0.338]\\
    GDP &  -0.058& [-0.062,0.054]  \\
    GINI & -0.020& [-0.020,-0.015]\\
    Market Penetration &   0.109& [0.104,0.114]\\
  \bottomrule
\end{tabular}
\end{table}
\begin{table}[H]
  \caption{Bootstrapped Coefficients From Multilevel Modeling for the effect of Tightness as a moderator on Dwell Time}
  \label{tab:tightness_appendix}
  \begin{tabular}{|c|c|c|}
    \toprule
    Fixed Effects& Estimate & $CI(95\%)$\\
    \midrule
    Intercept & 3.740 & [3.737,3.743]\\
    Strength of Ties & 0.116 & [0.111,0.121]\\
    Tightness & -0.061 & [-0.061, -0.060] \\
    Strength of Ties :  Tightness & 0.008 &  [0.006, 0.012]\\
    \hline
    \multicolumn{3}{|c|}{Control variables}  \\
    \hline
    Number of Friends &  -0.291& [-0.295,-0.285] \\
    GDP &  -0.156 & [-0.161,-0.150] \\
    GINI & -0.17 & [-0.171,-0.162] \\
    Market Penetration &  0.170& [0.169,0.170]\\
  \bottomrule
\end{tabular}
\end{table}

\section{Mixed effects model for the intersection of countries present across all three measures (for 1 run) }
\label{sec:culture_intersection}
\begin{table}[H]
  \caption{Coefficients from Multilevel Modeling for the effect of Individualism as a moderator on Dwell Time $(* p<0.05, ** p<0.01, *** p<0.001)$, Sample size: country = 18, dyads = 82800, RMSE = 2.947, AIC = 45373.1, BIC = 453824.6, R\textsuperscript{2} conditional =0.27, R\textsuperscript{2} marginal = 0.01}
  \label{tab:idv_appendix}
  \begin{tabular}{|c|c|c|}
    \toprule
    Fixed Effects& Estimate & Standard Error\\
    \midrule
    Intercept & 3.699*** &  0.126\\
    Strength of Ties & 0.147*** & 0.017\\
    Individualism & 0.085 & 0.116 \\
    Strength of Ties :  Individualism & -0.042*** & 0.011\\
    \hline
    \multicolumn{3}{|c|}{Control variables}\\
    \hline
    Number of Friends & -0.322***  &0.018 \\
    GDP &  -0.168* & 0.074\\
    GINI &  -0.171 & 0.063 \\
    Market Penetration & 0.238 & 0.128\\
  \bottomrule
\end{tabular}
\end{table}
\begin{table}[H]
  \caption{Coefficients From Multilevel Modeling for the effect of Mobility as a moderator on Dwell Time $(* p<0.05, ** p<0.01, *** p<0.001)$ Sample size: country = 18, dyads = 82800 RMSE = 3.07, AIC = 476936.1, BIC = 477029.3, R\textsuperscript{2} conditional = 0.09, R\textsuperscript{2} marginal = 0.01}
  \label{tab:mobility_appendix}
  \begin{tabular}{|c|c|c|}
    \toprule
    Fixed Effects& Estimate& Standard Error\\
    \midrule
    Intercept & 3.969*** & 0.123\\
    Strength of Ties & 0.126***& 0.014 \\
    High Mobility & 0.008 & 0.083\\
    Strength of Ties :  High Mobility & -0.037*** & 0.009 \\
    \hline
    \multicolumn{3}{|c|}{Control variables}\\
    \hline
    Number of Friends &  -0.30*** & 0.017 \\
    GDP & -0.152&  0.077\\
    GINI & -0.141 & 0.060\\
    Market Penetration & 0.099*** &0.010 \\
  \bottomrule
\end{tabular}
\end{table}
\begin{table}[H]
  \caption{Coefficients From Multilevel Modeling for the effect of Tightness as a moderator on Dwell Time $(* p<0.05, ** p<0.01, *** p<0.001)$, Sample size: country = 18, dyads = 82800, RMSE = 2.36, AIC = 482736.7, BIC = 482830, R\textsuperscript{2} conditional = 0.48, R\textsuperscript{2} marginal = 0.01 }
  \label{tab:tightness_appendix}
  \begin{tabular}{|c|c|c|}
    \toprule
    Fixed Effects& Estimate & Standard Error\\
    \midrule
    Intercept & 3.767*** & 0.135\\
    Strength of Ties &  0.164*** & 0.014\\
    Tightness & -0.021 & 0.084\\
    Strength of Ties :  Tightness & 0.094*** & 0.012 \\
    \hline
    \multicolumn{3}{|c|}{Control variables}\\
    \hline
    Number of Friends & -0.383*** & 0.024\\
    GDP & -0.183 & 0.081\\
    GINI & -0.180 & 0.070\\
    Market Penetration & 0.191 & 0.105\\
  \bottomrule
\end{tabular}
\end{table}

\onecolumn
\section{List of countries analyzed}
\label{sec:countrylist}
\begin{longtable}{|c|c|c|c|}
    \toprule
    Country & Individualism & Mobility & Tightness \\
    \midrule
    Argentina & 46 & $\times$ &$\times$\\
    Australia &90&4.308&4.4\\
    Austria &55&$\times$&6.8\\
    Belgium &75&$\times$ &5.6\\
    Brazil &38&4.419&3.5\\
    Canada &$\times$ &4.404&$\times$\\
    Chile &23&4.3&$\times$\\
    China & \multicolumn{3}{|c|}{excluded from analysis since Snapchat is banned} \\ 
    Colombia &13&4.483&$\times$\\
    Costa Rica &15&$\times$ &$\times$\\
    Czech Republic &58&$\times$ &$\times$\\
    Denmark &74&$\times$ &$\times$\\
    cEcuador &8&$\times$ &$\times$\\
    Egypt &38&3.971&$\times$\\
    El Salvador &19&$\times$ &$\times$\\
    Estonia &$\times$ &4.233&2.6\\
    Ethiopia &27&$\times$ &$\times$\\
    Finland &63&$\times$ &$\times$\\
    France &71&4.451&6.3\\
    Germany &67&4.194&7\\
    Ghana &20&$\times$ &$\times$\\
    Greece &35&$\times$ &3.9\\
    Guatemala &6&$\times$ &$\times$\\
    Hong Kong &25&4.043&6.3\\
    Hungary &55&3.893&2.9\\
    Iceland &$\times$ &$\times$ &6.4\\
    India &48&$\times$ &11\\
    Indonesia &14&$\times$ &$\times$\\
    Iran & \multicolumn{3}{|c|}{excluded from analysis since Snapchat is banned} \\
    Iraq &38&$\times$ &$\times$\\
    Ireland &70&$\times$ &$\times$\\
    Israel &54&4.336&3.1\\
    Italy &76&$\times$ &6.8\\
    Jamaica &39&$\times$ &$\times$\\
    Japan &46&3.934&8.6\\
    Jordan &$\times$ &3.96&$\times$\\
    Kenya &27&$\times$ &$\times$\\
    Kuwait &38&$\times$ &$\times$\\
    Lebanon &38&4.079&$\times$\\
    Libya &38&4.015&$\times$\\
    Malaysia &26&3.886&11.8\\
    Mauritius &$\times$ &4.385&$\times$\\
    Mexico &30&4.607&7.2\\
    Morocco &$\times$ &4.062&$\times$s\\
    Netherlands &80&4.448&3.3\\
    New Zealand &79&4.287&3.9\\
    Nigeria &20&$\times$ &$\times$\\
    Norway &69&$\times$ &9.5\\
    Pakistan &14&$\times$ &12.3\\
    Panama &11&$\times$ &$\times$\\
    Peru &16&$\times$ &$\times$\\
    Philippines &32&4.158&$\times$\\
    Poland &60&4.415&6.0\\
    Portugal &27&4.236&7.8\\
    Puerto Rico &$\times$ &4.603&$\times$\\
    Saudi Arabia &38&$\times$ &$\times$\\
    Sierra Leone &20&$\times$ &$\times$\\
    Singapore &20&4.133&10.4\\
    South Africa &65&$\times$ &$\times$\\
    South Korea &18&4.089&10.0\\
    Spain &51&4.415&5.4\\
    Sweden &71&4.364&$\times$\\
    Switzerland &68&$\times$ &$\times$\\
    Taiwan &17&4.118&$\times$\\
    Tanzania &27&$\times$ &$\times$\\
    Thailand &20&$\times$ &$\times$\\
    Trinidad and Tobago &$\times$ &4.421&$\times$\\
    Tunisia &$\times$ &3.954&$\times$\\
    Turkey &37&4.122&9.2\\ 
    Ukraine & \multicolumn{3}{|c|}{excluded from analysis due to geo-political instability} \\
    United Arab Emirates &38&$\times$ &$\times$\\
    United Kingdom &89&4.315&6.9\\
    United States &91&4.382&5.1\\
    Uruguay &36&$\times$ &$\times$\\
    Venezuela &12&4.508&3.7\\
    Zambia &27&$\times$ &$\times$\\
    \bottomrule
        \caption{List of Countries and the cultural values that they were surveyed for; $\times$ signifies country not surveyed for that cultural value}
    \label{tab:listcountry}
\end{longtable}
\twocolumn

\end{document}